\documentclass[namedreferences]{SolarPhysics}
\usepackage[optionalrh]{spr-sola-addons} 
\usepackage{graphicx}                    
\usepackage{color}                       
\usepackage{url}                         

\begin{document}
\begin{article}

\begin{opening}

\title{Automated Detection of Filaments and their Disappearance using Full Disc H$\alpha$ Images}

%
\author{A.~D.~Joshi{}$^{1}$\sep
        N.~Srivastava{}$^{1}$\sep
        S.~K.~Mathew{}$^{1}$      		
       }

%
\runningauthor{A.~D.~Joshi, N.~Srivastava and S.~K.~Mathew}
\runningtitle{Automated Detection of Filaments and their Disappearance}

%
  \institute{$^{1}$ Udaipur Solar Observatory
                     email: \url{janandd@prl.res.in} 
             }

\begin{abstract}

A new algorithm is developed that automatically detects filaments on the solar disc in H$\alpha$ images. Preprocessing of H$\alpha$ images include corrections for limb darkening and foreshortening. Further, by applying suitable intensity and size thresholds, filaments are extracted, while other solar features e.g. sunspots and plages are removed. Filaments attributes such as their position on the solar disc, total area, length, and number of fragments are determined. In addition, every filament is also labelled with a unique number. The algorithm is capable of following a particular filament through successive images which allows us to detect the changes and disappearance of the same, and thus provides a real-time warning of eruptive filaments. This aspect would prove to be of particular importance in studies pertaining to space weather. The algorithm will eventually be integrated with an upcoming telescope at Udaipur Solar Observatory for real time monitoring of activated/eruptive filaments.

\end{abstract}

%
\keywords{Prominences, dynamics; Prominences, quiescent}

\end{opening}

%

\section{Introduction}\label{sec:intro}
Filaments or prominences are low temperature and high density thread-like structures that are observed along the polarity reversal lines between regions of oppositely directed photospheric magnetic fields \cite{1998SoPh..182..107M}, suspended in the hot and tenuous corona, and supported by magnetic fields. Further, observations also show that there is always coronal magnetic field in the form of an arcade overlying the filament. Most of the filaments end their period of stay on the Sun by eruptions, where they rise slowly, and disappear over a few hours time. It is now well accepted that coronal mass ejections (CMEs) are very closely associated with such erupting filaments \cite{2002AdSpR..29.1451S,2003ApJ...586..562G}. It was also found by \inlinecite{2004JGRA..10910103S} that source regions of CMEs that are located close to the central meridian of the Sun are the most likely ones to give rise to major geomagnetic storms ($D_{st} \leq -100~\textrm{nT}$) at the Earth.

Therefore, the study of filaments assumes significance from the point of view of space weather. Other solar phenomena like CMEs (CDAW\footnote{SOHO LASCO CME Catalog at \url{http://cdaw.gsfc.nasa.gov/CME_list/index.html}}$\!$,  CACTUS\footnote{Computer Aided CME Tracking catalogue at \url{http://www.sidc.oma.be/cactus/}}$\!$, etc.), flares (TRACE Flare Catalog\footnote{TRACE Flare Catalog at \url{http://hea-www.harvard.edu/trace/flare_catalog/index.html}}) and sunspot groups (NOAA sunspot numbers\footnote{\url{http://www.solarmonitor.org/index.php}}) are very well catalogued, but at present a systematic cataloguing procedure for filaments is not in place. We realise that in order to achieve this, it is increasingly essential to have an automated system, and rely on human intervention as little as possible. With this in mind, we will present an algorithm that would not only detect a filament, but also its eruption in real-time.

In the past few years there have been many studies presenting an algorithm for filament detection. \inlinecite{2002SoPh..205...93G} have used a filament disappearance detection algorithm to conduct a statistical analysis of filaments for the year 1999 observed from the Big Bear Solar Observatory (BBSO). \inlinecite{2003SoPh..218...99S} (hereafter referred to as SK03) have employed advanced local thresholding technique and  morphological operations to efficiently detect filaments. \inlinecite{2005SoPh..228..119Q} on the other hand have made use of a nonlinear multiscale filtering technique, the stabilized inverse diffusion equation (SIDE), along with morphological operations to detect filaments and their disappearance. \inlinecite{2005SoPh..228...97B} have also developed an algorithm that detects filaments, and characterises its spine, direction of barbs, and hence determines the chirality. \inlinecite{2005SoPh..227...61F} have presented filament detection method, which also uses various image segmentation techniques for H$\alpha$ spectroheliograms.

What we present in this paper, is a new filament extraction and tracking algorithm, which can be used during real-time observations. We have developed this algorithm as a substitute to the morphological operations in order to suit our requirements. Eventually, we would like to integrate the whole procedure with a new instrument, the Dual Beam H$\alpha$ Doppler System, that is under development at USO. The instrument has been planned to observe the full disc of the Sun in two beams; a low-cadence beam (\emph{mode a}) to monitor the activation of filaments, and a high-cadence beam (\emph{mode b}) that would record Doppler shifts at several positions along the H$\alpha$ line profile for activated erupting filaments. Using this, we can obtain the true velocities of erupting filaments. We plan to use the algorithm presented here for \emph{mode a} to detect an activated filament, which would provide a real-time signal of a filament that is going to erupt. Based on this signal a threshold for activated filaments would be decided, on crossing which \emph{mode b} would start taking the high-cadence images through a tunable Fabry-Perot filter, along the H$\alpha$ line profile.

\section{Data reduction}\label{sec:reduce}
We have used full-disc H$\alpha$ images from Udaipur Solar Observatory (USO), taken on 2008 Apr 26 and from Kanzhelh$\ddot{o}$he Solar Observatory (KSO) taken on 2005 Jan 05 for this study. There is a quiescent filament close to the disc centre on 2008 Apr 26, parts of which erupted later during the observing period. On 2005 Jan 05, there is a large quiescent filament stretching for more than half a solar radius in the North-Western part of the solar disc. This filament completely erupted in a very short period of time.

\subsection{Preprocessing}
This step involves dark subtraction and flat fielding of the images. Since different images have different intensities, it is necessary to bring them to a certain defined level, after which comparison between them can be handled in a convenient and uniform way. Hence, the normalisation procedure is used. The formula used for the same is:
\begin{equation}\label{eq:limbda}
I_{norm} = (I_{in}-c)\times\frac{b-a}{d-c} ~+~ a,
\end{equation}
where, $I_{in}$ and $I_{norm}$ are the input and output image intensities respectively; $a$ and $b$ are the desired minimum and maximum intensities of the output image; and $c$ and $d$ are the minimum and maximum intensities of the input image.

In solar images, it is very much possible that there are some `outliers', i.e. points where the intensity is either very much less than or very much more than the average intensity in the image. If we simply have these intensities as the values for $c$ and $d$ in the equation above, we will get an output image with very low contrast. To avoid this, $c$ and $d$ are taken to be the minimum and maximum intensities of the input image after removing 5\% pixels having the lowest intensity and 5\% pixels having the highest intensity.

\begin{figure} 
\centerline{\includegraphics[width=0.75\textwidth]{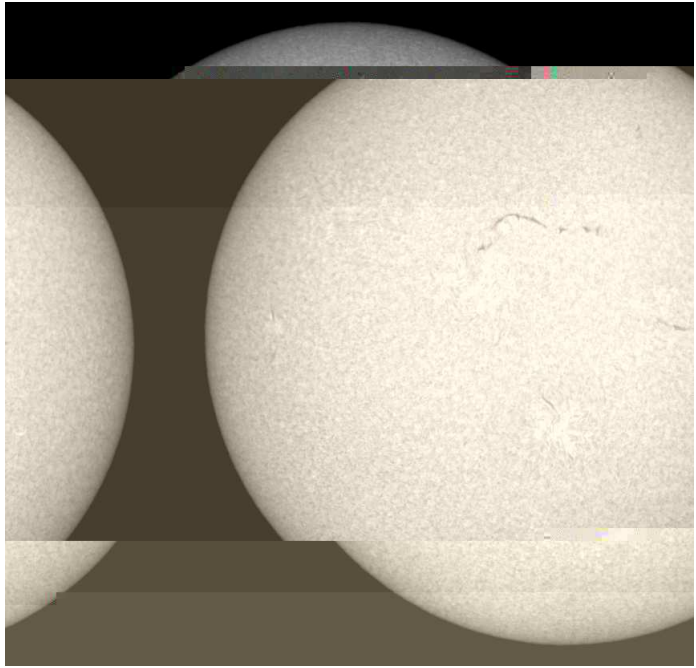}}
\caption{Full disc H$\alpha$ image taken from Udaipur Solar Observatory on 2008 Apr 26 after applying dark subtraction, flat fielding and normalisation.}\label{fig:preproc}
\end{figure}

\subsection{Solar disc alignment}
For the series of consecutive images, the solar disc may not always be at a fixed position in all of them, because of the errors in telescope tracking. In order to analyse the images in sequence, it is necessary to bring the solar disc in each of them at a fixed position, preferably at the centre of the image. To achieve this, we have developed a program that  detects the radius and coordinates of centre of the solar disc in each image. For this, derivatives along several rows and columns were computed. The point at which the derivative was minimum or maximum was a point on the edge of the disc, because of the steep intensity gradient. Thus, the centre and radius of each image was determined and the image was shifted so as to bring the solar disc at its centre. This method is a lot simpler and time-saving compared to the manual selection of a feature and then shifting the image.

\subsection{Limb darkening removal}
In the case of photospheric and chromosphere full disc images, limb darkening is a major issue \cite{1977SoPh...51...25P,1994SoPh..153...91N}. As light emitted close to the limb passes through a thicker slab of atmosphere compared to the light coming from the disc centre, we see the limb darker than the disc centre. To correct for this effect, the average centre to limb variation (CLV) was determined for each image by finding the CLV in intervals of 0$.5^{\circ}$. A 5$^{\mathrm{th}}$ order polynomial was fitted to average CLV, and a full disc image was created using this polynomial function with the same radius and centre coordinates as the input image. The normalised input image was then divided by this image and multiplied by the average value of the disc intensity to provide the image with limb darkening removed.

\begin{figure}
\centerline{\includegraphics[width=0.75\textwidth]{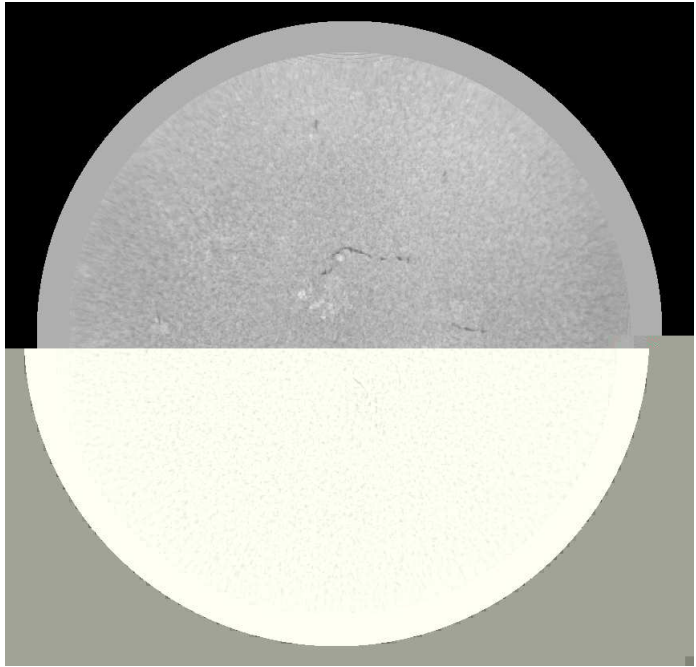}}
\caption{Image in Figure~\ref{fig:preproc} after applying corrections for limb darkening and foreshortening. The region greater than $r \geq 0.9$ is filled with the average value at the centre, to cover the gaps in the image after applying Eqns~\ref{eq:foreshort} and \ref{eq:forearea}.}\label{fig:limbfore}
\end{figure}

\subsection{Foreshortening correction}
Since the Sun is a spherical body, an image of the disc of the Sun suffers from `foreshortening', i.e., an area of, say, 10\,$\times$\,10 pixels on the disc covers a larger area on the Sun if it is present close to the limb than if it is present near the disc centre. As a consequence, a solar feature present near the solar limb would appear smaller than its actual size on the Sun. The correction applied to the images is given below (SK03).
\begin{equation}\label{eq:foreshort}
r' = \,\sqrt{1 - \sqrt{\,1 - r^2}},
\end{equation}
where, $r$ is the fractional radius of solar disc in the original image and $r'$ is the fractional radius of solar disc in the corrected image.

However, it is not enough to simply map intensity at a distance $r$ to a distance $r'$, because the area around the point $r$ would not be preserved in this case. To preserve the area, the following area transformation given by  \inlinecite{1996GSur..178.....E},  \inlinecite{1988JApA....9..137A}.
\begin{equation}\label{eq:forearea}
A_{corrected} = \frac{A_{apparent}}{0.2\,r + \sqrt{\,1 - r^{2}}}
\end{equation}
Here, as before, $r$ is the fractional solar radius. Thus, when the correction is applied, an area $A_{apparent}$ located at a distance $r$ is rebinned to area $A_{corrected}$ and placed at a distance $r'$.

From Eqn.~\ref{eq:forearea} we can see that as we go closer and closer to the limb ($r\rightarrow1$), $A_{corrected}$ goes as $r^{-1}$, and at $r = 1$, $A_{corrected}$ grows to five times $A_{apparent}$. In such a case, even after rebinning, there are a few on the corrected solar disc, which cannot be mapped onto. To counter this issue, we have carried out the correction only up to $r = 0.9$. This is justifiable, since, no filament of our current interest is located so close to the solar limb.

\section{Filament extraction}\label{sec:filex}
To extract solar filaments from H$\alpha$ images, SK03 as well as \inlinecite{2005SoPh..227...61F} have made use of morphological image processing tools, namely closing and opening. On using the same technique, it was observed that area of the filament was not preserved. A filament may exist in the form of a number of broken fragments, especially when it is activated. The broken fragments were found to merge on application of morphological closing. To be able to detect activated or disappearing filaments, it is advisable to keep the filament area as accurate as possible. Therefore, in our study, we have used a different method, a rigorous algorithm where each contiguous group of pixels is checked to decide if it is a filament fragment.

\subsection{Intensity and size threshold}

This process involves comparing intensity of each pixel of the image with an \textit{intensity threshold} to determine if the pixel is part of the filament (1 or white) or if it is part of the background (0 or black). If we use a uniform value of threshold, it is known as a \textit{global} threshold, while if we use threshold that changes all over the image, it is known as a \textit{variable} threshold.
\begin{equation}\label{eq:thresh0}
g(x,y) = 
\left\{
\begin{array}{rl}
\vspace{0.1cm}
1 \quad & \textrm{for}\;f(x,y) \: > \: T_{xy}\\
0 \quad & \textrm{for}\;f(x,y) \: \leq \: T_{xy}
\end{array}
\right.
\end{equation}
The important thing to do now is find the threshold function in Eq~\ref{eq:thresh0}. We have used the variable local thresholding method described in \inlinecite{2002dip..book.....G} and SK03 for this purpose. The first step is calculation of median at every pixel in the image over a 19\,$\times$\,19 neighbourhood centred on the pixel, denoted as $med_{19}(x,y)$. A lower cutoff value ($I_{inf}$) and a higher cutoff value ($I_{sup}$) are selected, which are respectively 10\% and 90\% of the total intensity range of the image. If the median value at a pixel is less than $I_{inf}$, threshold is equal to binary value 0, and if it is greater than $I_{sup}$, threshold is eauql to binary value 1. If the pixel intensity lies in between $I_{inf}$ and $I_{sup}$, threshold is equal to the median value calculated in the previous step (SK03).
%
\begin{equation}\label{eq:thresh1}
T_{xy} = 
\left\{
\begin{array}{rcl}
\vspace{0.1cm}
I_{inf} \quad & \textrm{for} \; & med_{19}(x,y) \: \leq \: I_{inf}\\
\vspace{0.1cm}
med_{19}(x,y) \quad & \textrm{for} \; & I_{inf} \: \leq \: med_{19}(x,y) \: \leq I_{sup}\\
I_{sup} \quad & \textrm{for} \; & med_{19}(x,y) \: \geq \: I_{inf}
\end{array}
\right.
\end{equation}

This gives a binary image in which very bright regions are marked as white, and very dark regions, which also includes area of the image outside the solar disc are marked as black; while the rest of the image is marked black or white based on the threshold calculated in the formula above.

\begin{figure}
\centerline{\includegraphics[width=0.75\textwidth]{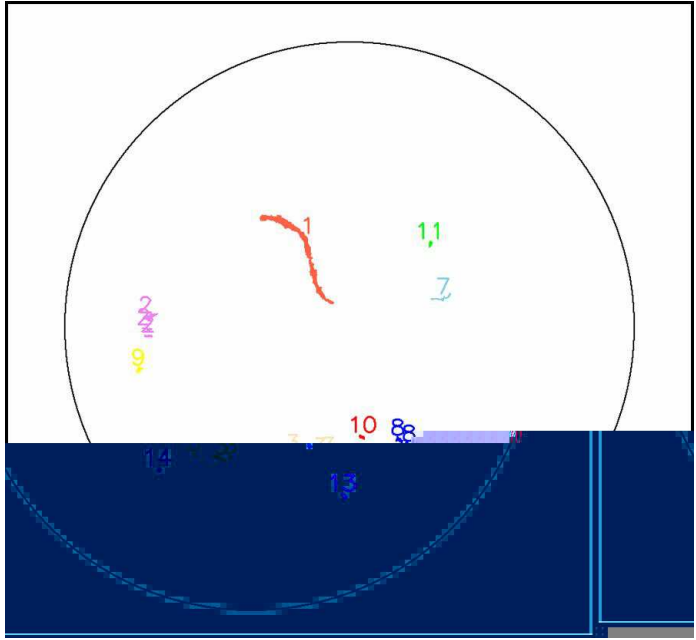}}
\caption{Binary image obtained after applying intensity and size  thresholds for the KSO image on 2005 Jan 05. The black circle marks the edge of the solar disc. Fragments belonging to the same filament are colour-coded and labelled.}\label{fig:colbase}
\end{figure}

However, the image also contains some features which when compared with the original image, are seen to be not filaments, but regions where intensity is similar to that of filaments, like the granular pattern in chromosphere. But we found that these regions are typically less than 12 pixels in size. Hence a size threshold is applied where contiguous white regions less than 12 pixels in size are turned to black. At the end of this step, we get a binary image, as shown in Figure~\ref{fig:colbase}, with all the filaments identified as differently coloured regions. The black circle shown is the radius of the solar disc.

\subsection{Filament identification}

It is observed that a single large filament is broken into several small fragments. The image obtained after thresholding also contains several such regions. We employ the \emph{40-pixel-distance criterion} to identify fragments of a single filament \cite{2002SoPh..205...93G}. First of all, pixel coordinates of all the contiguous regions are extracted. The region with the largest fragment is labelled \emph{1}, the next largest is labelled \emph{2}, and so on. The fragment labelled \emph{1} is then compared with all other fragments, to check if the two lie within a distance 40 pixels from each other. Assuming that filaments are thread-like structures, this step takes care of the possibility that the two fragments could lie one after the other in any direction, and in the rare case of they lying side by side. Any fragment thus identified lying close to fragment lebelled \emph{1} is also labelled \emph{1}. Such a fragment is again compared with all remaining fragments to check if it lies within a distance of 40 pixels. Once fragment \emph{1} is checked with all other fragments, fragment \emph{2} is checked with the remaining fragments. This procedure is repeated till we run out of all the fragments.

%


\begin{figure}

\centerline{
	\hspace*{0.015\textwidth}
	\includegraphics[width=0.495\textwidth,clip=]{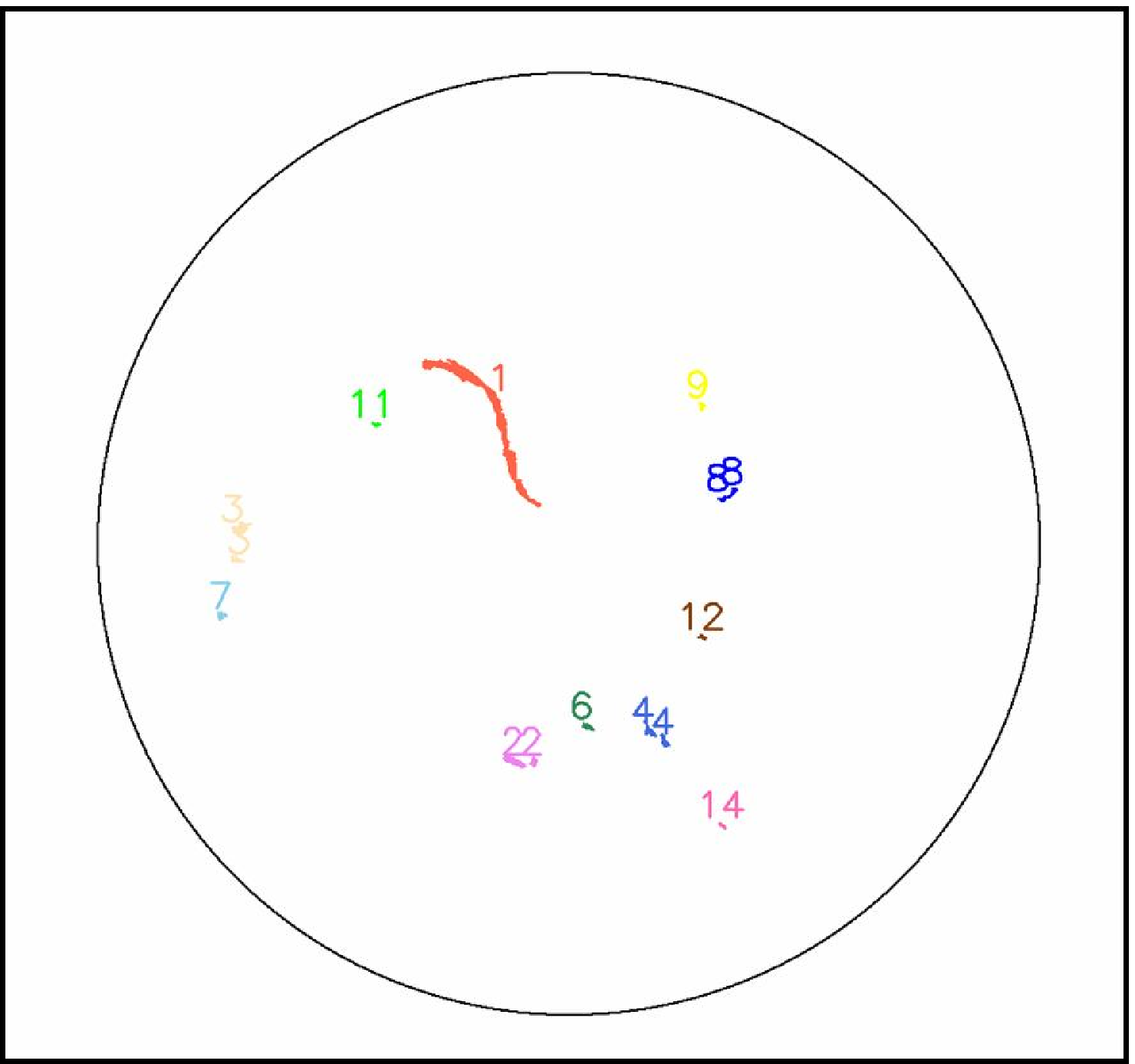}
    \hspace*{-0.01\textwidth}
    \includegraphics[width=0.495\textwidth,clip=]{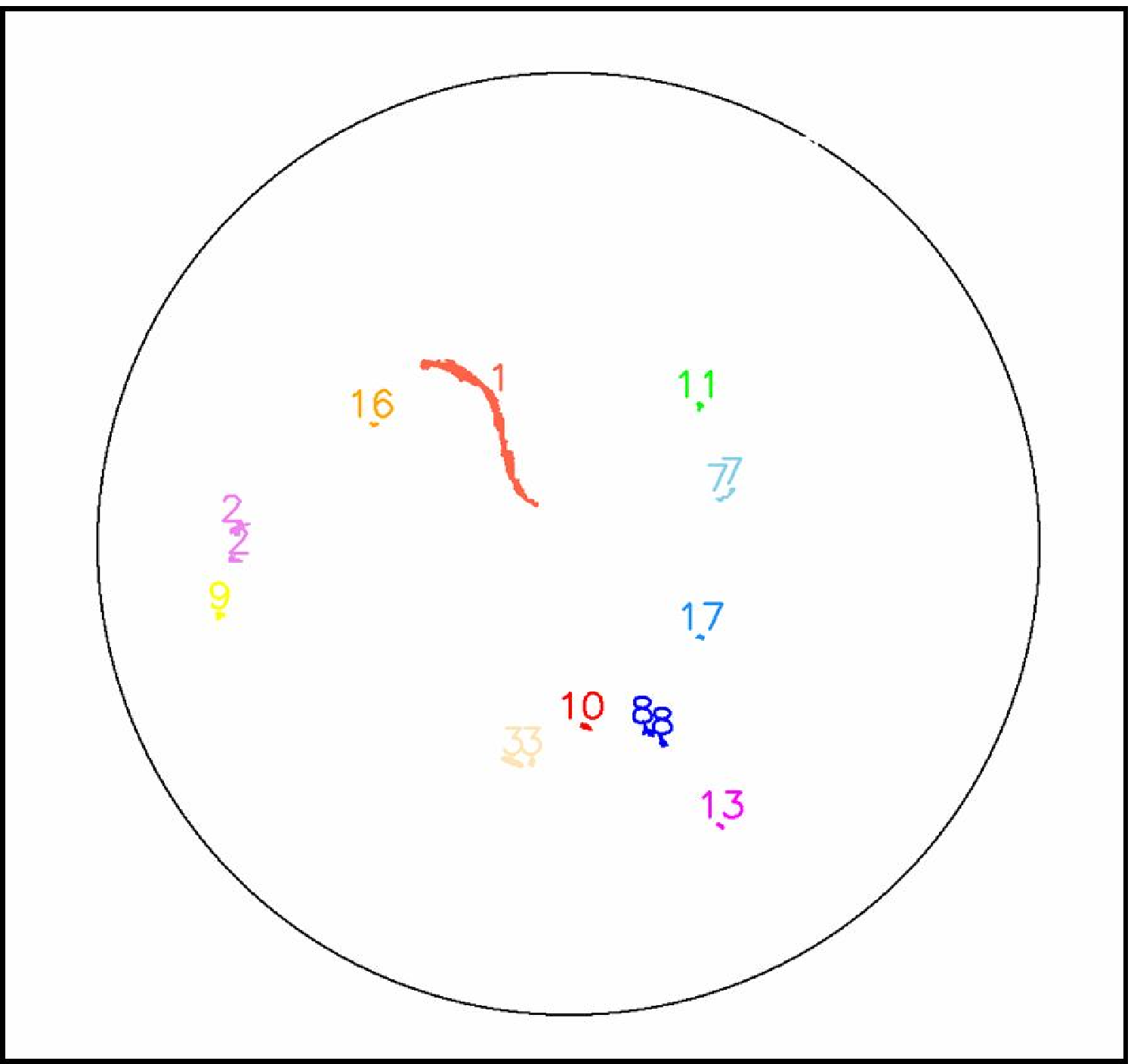}
          }
\vspace{-0.34\textwidth}
\centerline{
	\Large \bf
	\hspace{0.0\textwidth}   
    \hspace{0.44\textwidth}  
    \hfill
    	    }
\vspace{0.31\textwidth}

\caption{\textit{Left panel:} As compared to Figure~\ref{fig:colbase}, labelling in the a consequent image has gone wrong because of non-uniform fragment sizes. \textit{Right panel:} The corrected labelling after employing the 15-pixel-distance criterion. Also note that newly detected fragments are assigned new labels.}\label{fig:colf2}
\end{figure}

In order to track the filament across all the images in a given day, it is necessary to retain the labels. For the next image in the sequence, the filament identification step remains the same, at the end of which, we would get appropriate fragments of the next image grouped into a single filament, and identified with a unique label. However, as fragments may change their shape as well as size with time, and labelling of fragments depends on their size, the labelling for the next image may be different from that in the previous image. Hence, we have used the  \emph{15-pixel-distance criterion}, where filaments of the new image are compared with filaments of the previous image, and if they are found to lie at a distance of less than 15 pixels from each other, filament label of the original image is assigned to the corresponding filament of the new image. Taking into account the new fragments coming up in images, or multiple fragments merging to form a single fragment, the 15-pixel-distance criterion proved to be a satisfactory method to keep track of a given filament.



\begin{figure}

\centerline{
	\hspace*{0.015\textwidth}
	\includegraphics[width=0.495\textwidth,clip=]{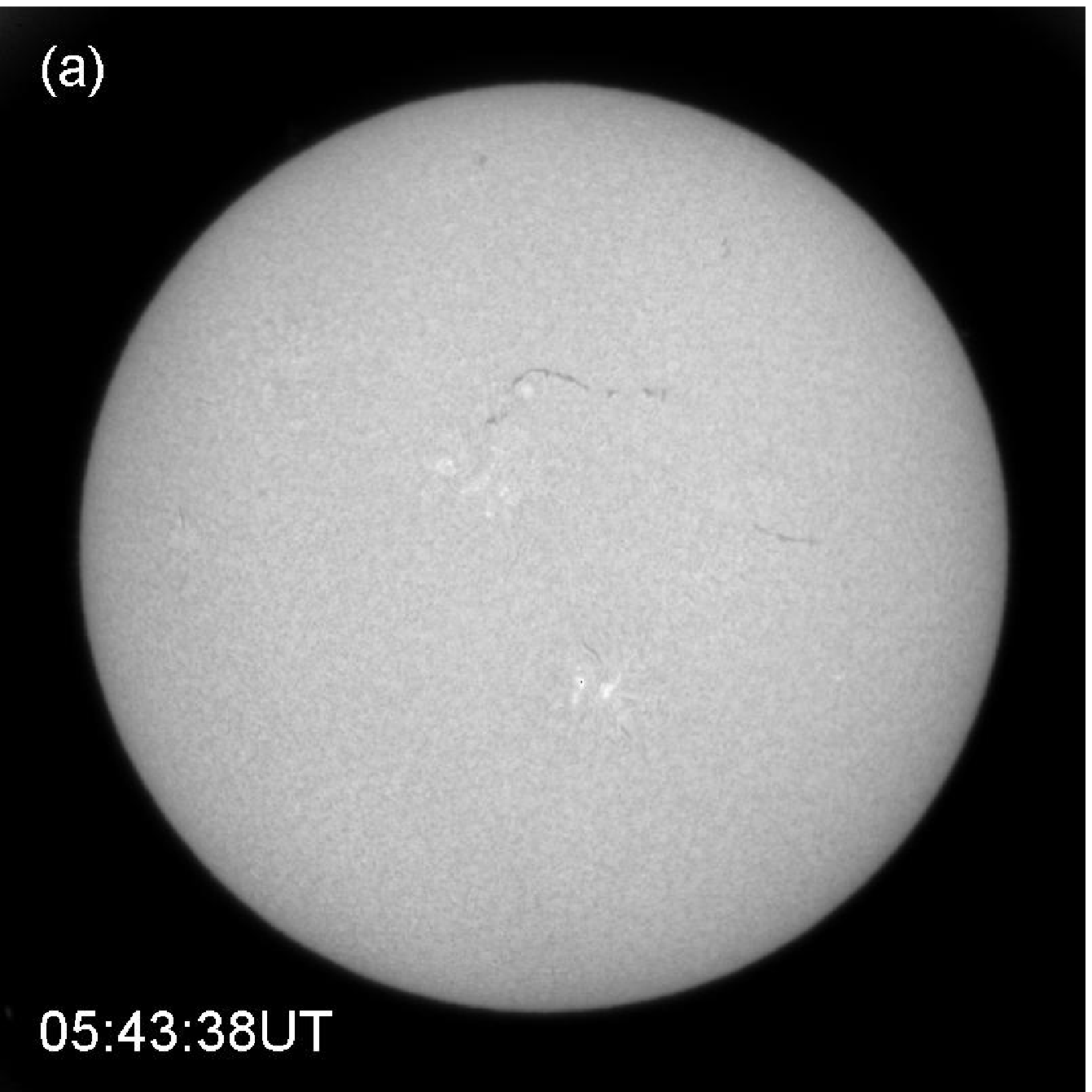}
    \hspace*{-0.01\textwidth}
    \includegraphics[width=0.495\textwidth,clip=]{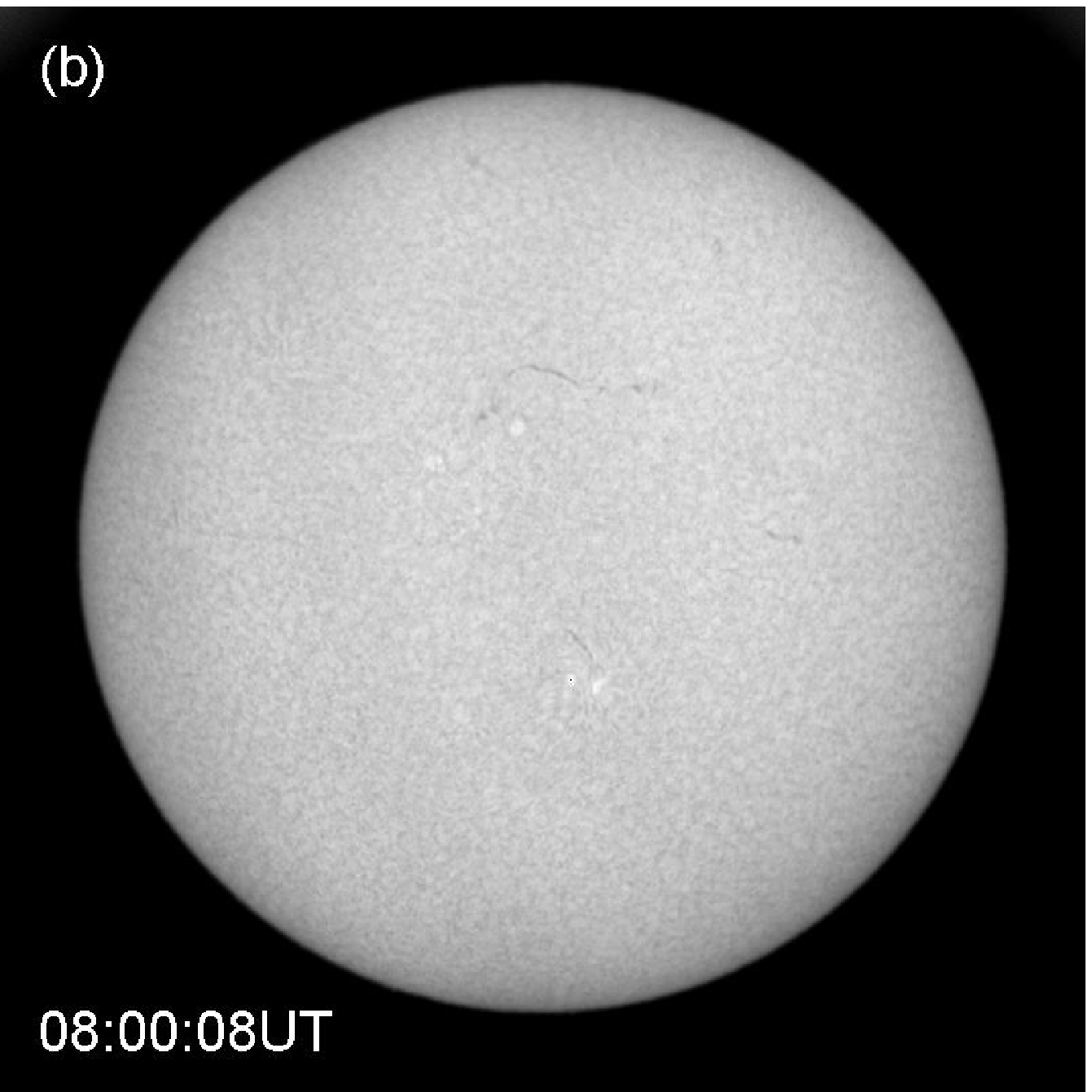}
          }
\vspace{-0.34\textwidth}
\centerline{
	\Large \bf
	\hspace{0.0\textwidth}   
    \hspace{0.44\textwidth}  
    \hfill
    	    }
\vspace{0.31\textwidth}

\caption{H$\alpha$ images from Udaipur Solar Observatory, for 2008 Jan 26 showing the filament in its stable state (\textit{a}) and while it is disappearing (\textit{b}).}\label{fig:mos-26apr}
\end{figure}

\begin{figure}

\centerline{
	\hspace*{0.015\textwidth}
	\includegraphics[width=0.495\textwidth,clip=]{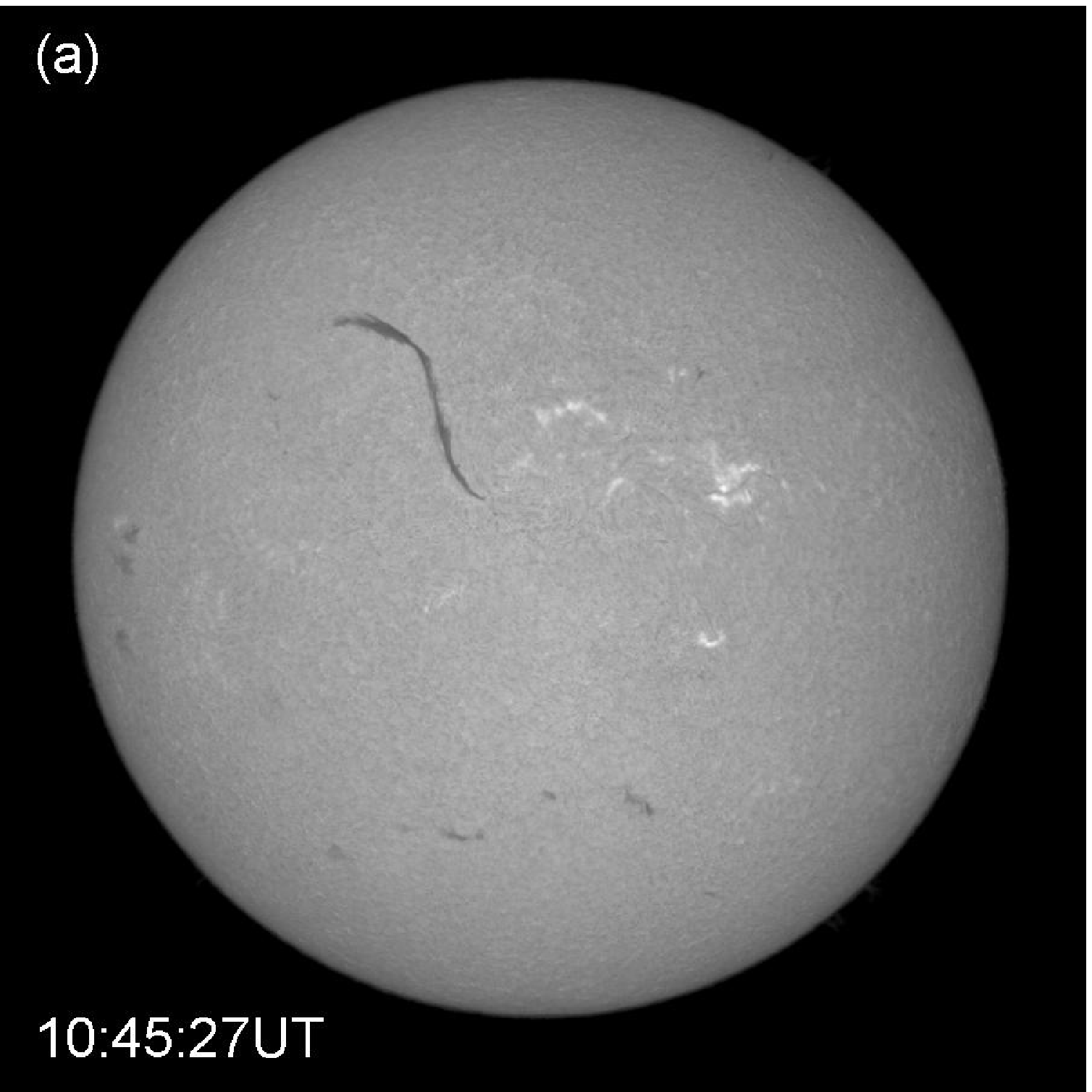}
    \hspace*{-0.01\textwidth}
    \includegraphics[width=0.495\textwidth,clip=]{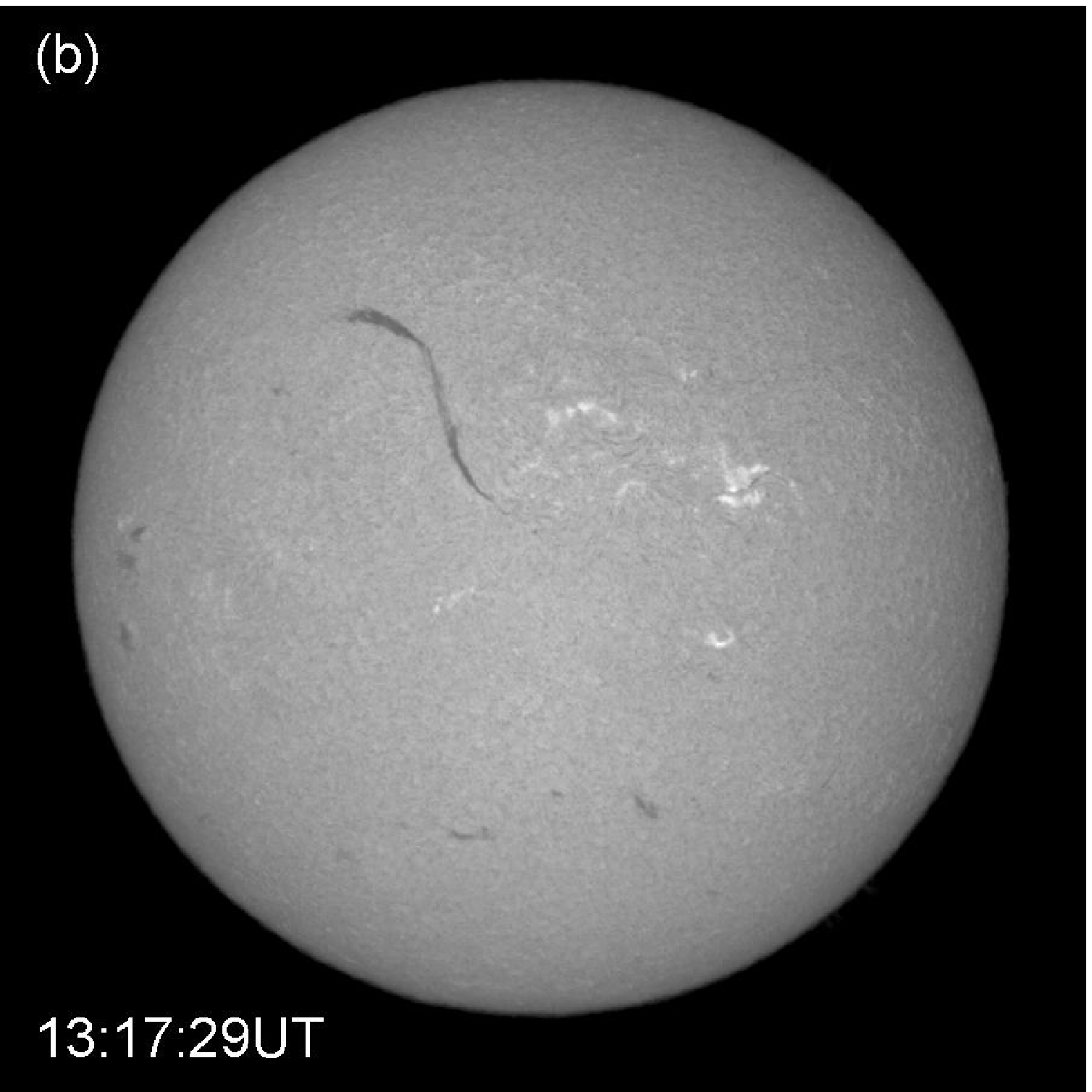}
          }
\vspace{-0.34\textwidth}
\centerline{
	\Large \bf
	\hspace{0.0\textwidth}   
    \hspace{0.44\textwidth}  
    \hfill
    	    }
\vspace{0.31\textwidth}

\centerline{
	\hspace*{0.015\textwidth}
    \includegraphics[width=0.495\textwidth,clip=]{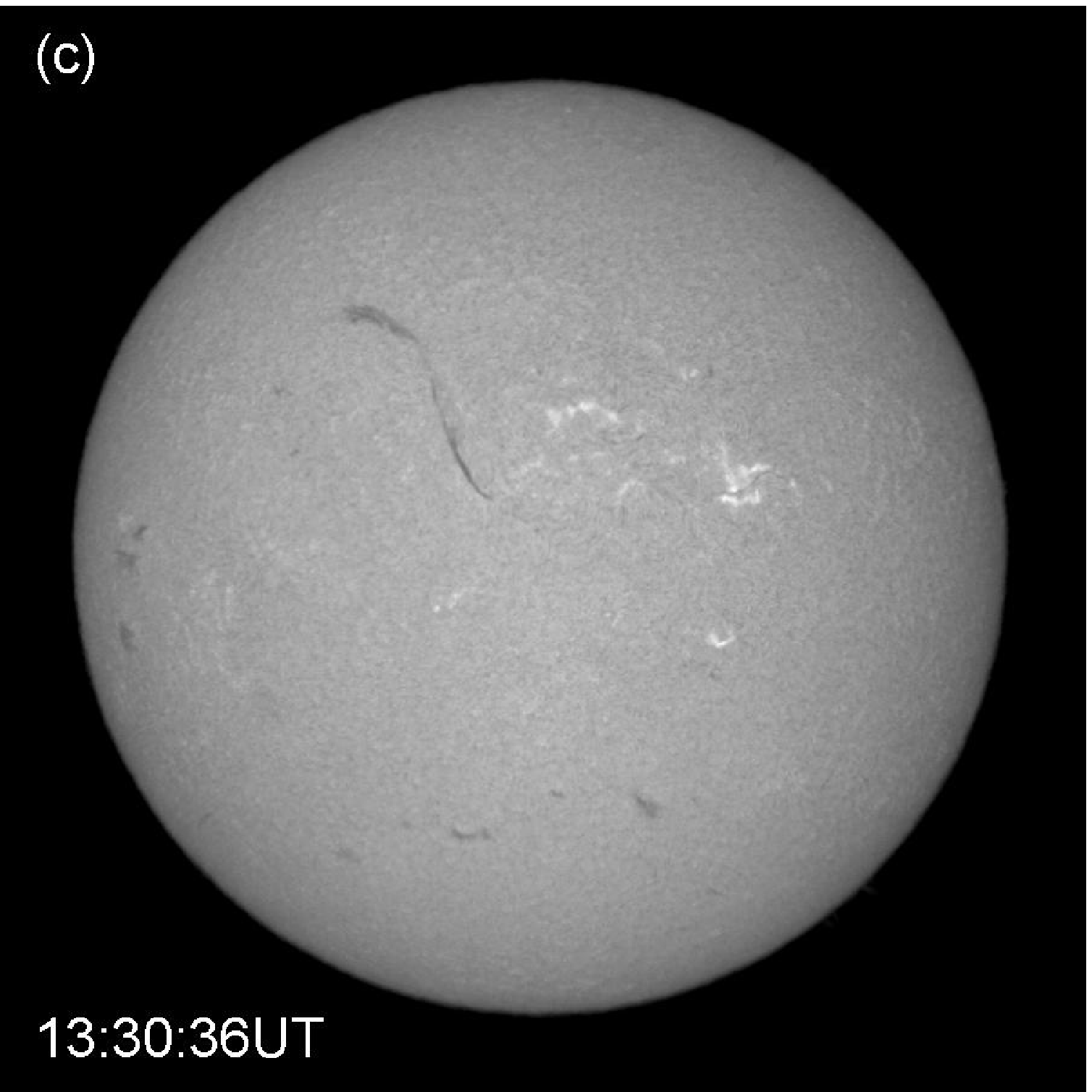}
    \hspace*{-0.01\textwidth}
    \includegraphics[width=0.495\textwidth,clip=]{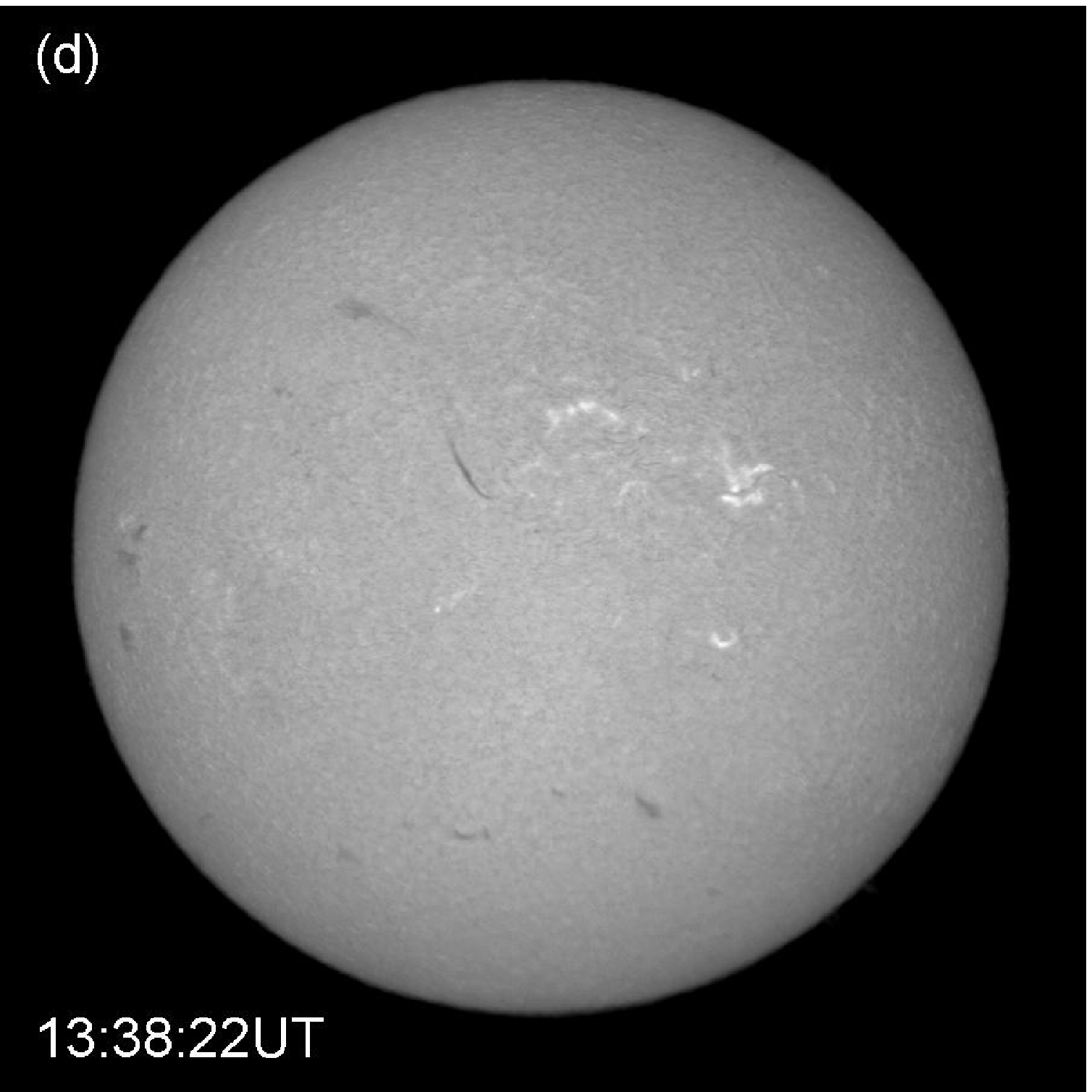}
    	   }
\vspace{-0.35\textwidth}
\centerline{
	\Large \bf
	\hspace{0.0\textwidth}	
    \hspace{0.44\textwidth}	
    \hfill
    	   }
\vspace{0.31\textwidth}

\caption{H$\alpha$ images from Kanzelh$\ddot{o}$he Solar Observatory, for 2005 Jan 05 showing the filament and its rapid eruption at different times.}\label{fig:mos-05jan}
\end{figure}

\section{Results and discussion}\label{sec:rnd}
\begin{figure}
\centering
\includegraphics[width=0.75\textwidth]{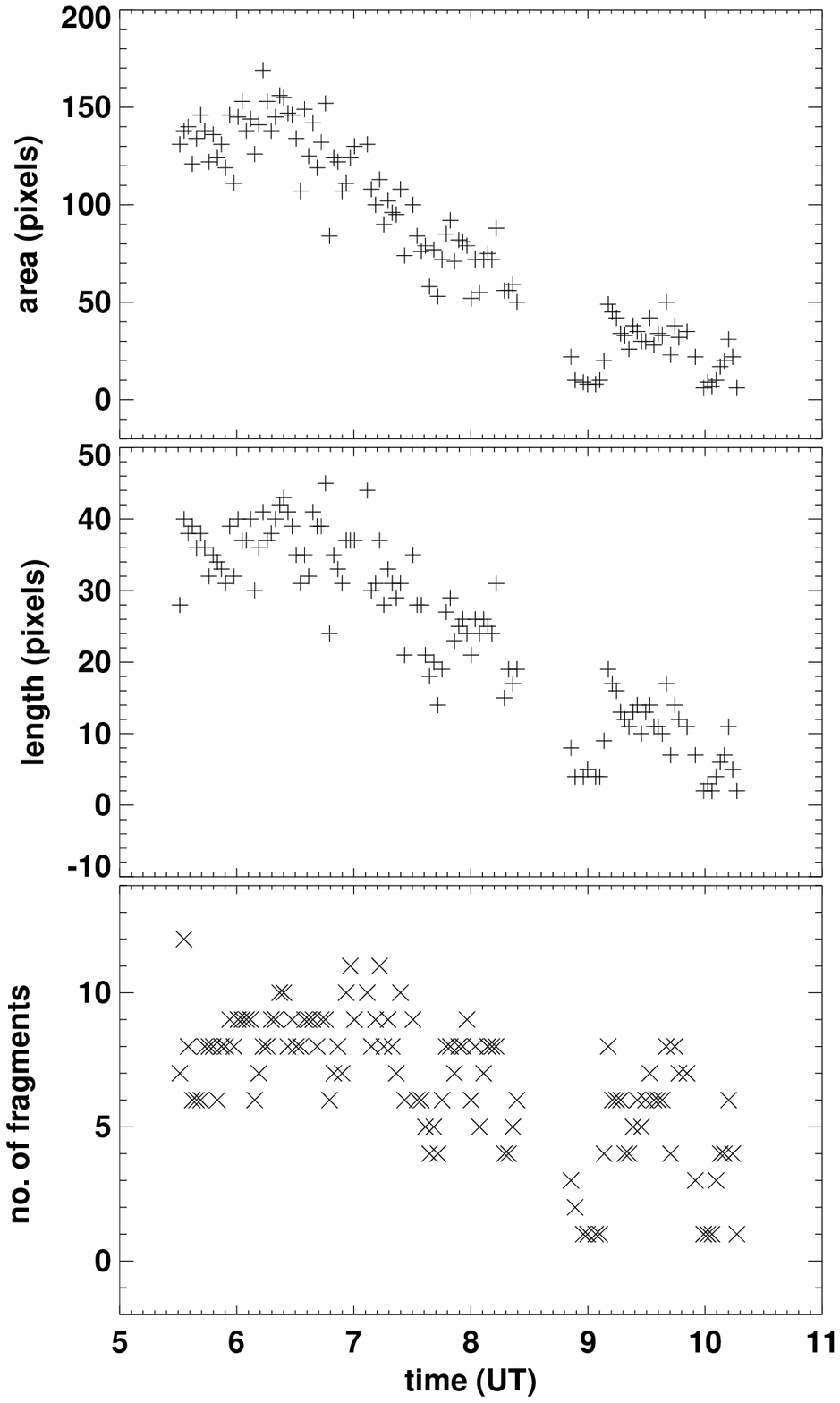}
\caption{Total area (\textit{top panel}), total length (\textit{middle panel}), and number of fragments (\textit{bottom panel}) of the filament labelled~\emph{1} observed on 2008 Apr 26.}\label{fig:res-26apr}
\end{figure}

\begin{figure}
\centering
\includegraphics[width=0.75\textwidth]{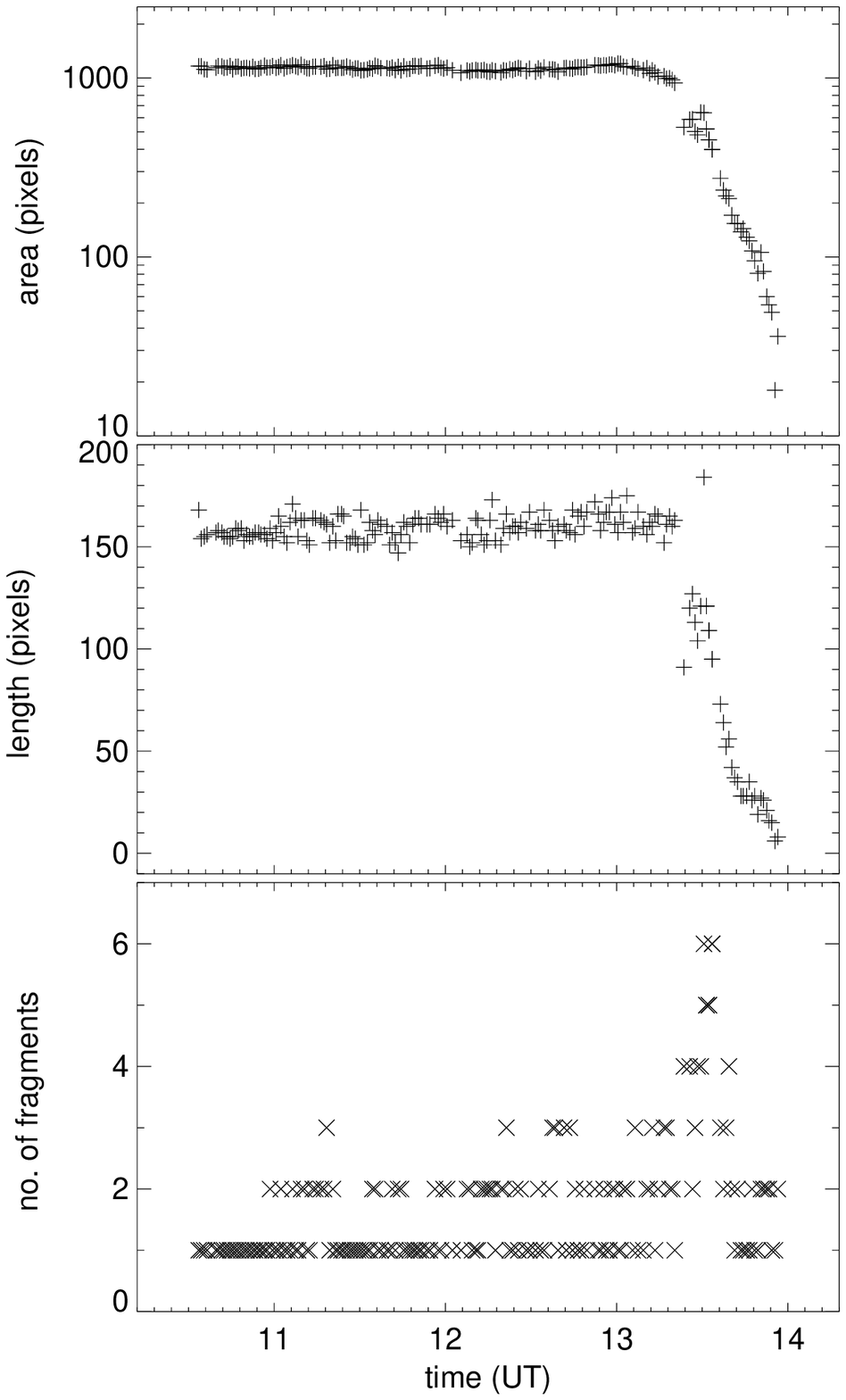}
\caption{Total area (\textit{top panel}), total length (\textit{middle panel}), and number of fragments (\textit{bottom panel}) of the filament labelled~\emph{1} observed on 2005 Jan 05.}\label{fig:res-05jan}
\end{figure}

The program was developed in Interactive Data Language (IDL), version 6.3 at USO, and was used to track and label the filaments. The program was used to analyse 132 images for 2008 Apr 26 and 178 images for 2005 Jan 05. Figures~\ref{fig:mos-26apr} and \ref{fig:mos-05jan} shows images for the two events of filament disappearance studied. The program computes total area of a given filament in terms of pixels, the sum of lengths of all the fragments identified to be belonging to the filament, and the number of fragments making up the filament. While a binary image obtained after thresholding is being analysed by the program, we have information about position of all points of the filaments present in it, which can be used to find the intensity changes in the image.
The results for the two filaments detected, are shown in Figures~\ref{fig:res-26apr} and \ref{fig:res-05jan}.

We find from Figure~\ref{fig:res-26apr} that area as well as length of filament \emph{1} (the filament close to the disc centre in Figure~\ref{fig:mos-26apr}) decreases gradually over time from around more than 150 pixels to about 20 pixels. Indeed images of the full day show the filament gradually disappearing. However, the algorithm could not detect the filament for the full duration for which observations were carried out. This was primarily due to the deteriorating seeing conditions at the observing site; it becomes difficult to observe the same filament even without bringing the algorithm into picture.

Figure~\ref{fig:mos-05jan} shows the filaments on 2005 Jan 05 at different times. The long thick filament, labelled \emph{1}, is seen to erupt very rapidly in a time span of less than an hour. Figure~\ref{fig:res-05jan} shows the total area, total length and the number of fragments of the same filament. We can see here that the area and length of this filament remains almost constant for most of the time. It is only over the last 50 minutes that the filament disappears. From the figure we can also say for sure that the disappearance began at 13:06 UT. The number of fragments of this particular filament does not change much for as long as the filament area is constant. However, as the filament starts to disappear, we can see it breaking up into several fragments. However, this number goes down as rapidly as it rises and, at the end there are only a few small fragments left behind (bottom panel in Figure~\ref{fig:res-05jan}).

To carry out corrections for limb darkening, foreshortening, and applying the algorithm, a regular desktop computer required on an average 1 minute for one image. We would be operating \emph{mode a} of the Dual Beam H$\alpha$ Doppler System at a cadence of around 1 image per minute. Thus the algorithm is well-suited to be operated in real-time for giving out a warning of a possible filament eruption.

It should be noted that the algorithm was built keeping in my mind our primary interest---the disappearance of filaments. It is observed that during the disappearance, parts of the filament lose their material. Some of the fragments totally disappear, while some others shrink in size. In this regard, we found that our algorithm fares better than the technique given in SK03, since we do not lose information on any of the fragments constituting a filament, nor do we fill the gaps between two neighbouring fragments.


\section{Summary}\label{sec:summ}
This algorithm is capable of routinely cataloguing filaments including the eruptives ones, with their attributes such as location, length, area and number of fragments. Another important information that can be sought using this algorithm is to establish a criterion for eruptive filaments based on a sudden or gradual change in its length, area and the number of fragments, hours prior to it complete eruption.

Using this algorithm with the full disk H$\alpha$ images on a near real-time basis will help us in understanding the process of eruption of the filament itself. Recently, \inlinecite{2007BASI...35..447J}  have shown that slow ascent of filaments is often observed in many eruptive prominences that are associated with CMEs. In fact from an analysis of He 304~$\textrm{\AA}$ images it was found that this slow ascent can be detected in 75\% of quiescent eruptive filaments. As the time duration for this slow ascent varies between 1-4 days prior to their eruption, the detection of ascent will be crucial. This can be used for forecasting of eruptive filaments and hence their associated CMEs, with very few false alarms.

Recently \inlinecite{2008AnGeo..26.3061M} presented a broad concept for the build-up to eruptive events which consists of a CME, an eruptive prominence, a cavity around the filament and a flare. This is primarily based on the hypothesis that CMEs are causally linked to the formation, evolution and maintenance of filament channels and filaments. They suggest various stages in the build-up to eruption. The algorithm is capable of detecting, extracting and estimating various attributes of filaments with better accuracy and will thus be used for testing the concept of how CMEs are linked to the evolution of filament channel, filament evolution and associated magnetic fields.

%

%

%

%
\begin{acks}
The authors gratefully acknowledge Wolfgang Otruba for providing full-disc H$\alpha$ images from KSO, courtesy of the \emph{Central European Solar ARchives} (CESAR). The authors acknowledge the help provided by Engineering Trainees at USO, Nisarg K.~Bhatt and Shivam M.~Desai in developing parts of the program for filament extraction. We also thank Gabrudeen Khan, USO, for observing the event analysed in this paper. Work by one of the authors (NS) contributes to the research on collaborative NSF grant ATM-0837915 to Helio Research.
\end{acks}

%
%
 \bibliographystyle{spr-mp-sola-cnd} 
 \bibliography{aufil_det}
%
%
%
%

\end{article} 

\end{document}